\documentclass[fdp,fleqn]{w-art}
\usepackage{times}
\usepackage{w-thm}
\usepackage{amssymb}
\usepackage{bbm}
\usepackage[]{graphicx}

\begin{document}
\pagespan{1}{}

\title[An involuted orbifold MSSM]{An involuted orbifold MSSM}

\author[P.K.S. Vaudrevange]{Patrick K. S. Vaudrevange\inst{1,}%
  \footnote{Author E-mail:~\textsf{Patrick.Vaudrevange@physik.uni-muenchen.de}}}
\address[\inst{1}]{Arnold Sommerfeld Center for Theoretical Physics, Ludwig-Maximilians-Universit\"at M\"unchen, 80333 M\"unchen, Germany}

\begin{abstract}

A compactification of the $\text{E}_8\times\text{E}_8$ heterotic string on a $\mathbbm{Z}_2\times\mathbbm{Z}_2$ orbifold equipped with an additional freely acting involution is presented. This model reproduces the exact chiral MSSM spectrum with matter parity and a non-trivial Yukawa structure. The key ingredient is a freely acting Wilson line associated to the involution, breaking $\text{SU}(5)$ to $\text{SU}(3)\times\text{SU}(2)\times\text{U}(1)_Y$.
This work is based on a talk given at the ``9th Hellenic School and Workshop on Elementary Particle Physics and Gravity'' and reviews the results of a collaboration with M.~Blaszczyk, S.~Groot Nibbelink, M.~Ratz, F.~R\"uhle and M.~Trapletti.

\end{abstract}

\maketitle

\section{Introduction}
Compactifications of the heterotic string on orbifolds~\cite{Dixon:1985jw,Dixon:1986jc} yield, since the early days, a promising framework for studying phenomenological aspects of string theory, see e.g.~\cite{Ibanez:1986tp,Ibanez:1987sn} or~\cite{Kobayashi:2004ya,Buchmuller:2005jr,Buchmuller:2006ik,Lebedev:2006kn,Lebedev:2007hv} for later work and \cite{Nilles:2008gq} for a recent review. The world-sheet theory of strings on orbifolds is described by a combination of free conformal field theories (CFT) and hence allows for explicit string calculations. This is in contrast to a general smooth  Calabi-Yau (CY) compactification, where the world-sheet theory involves a complicated interacting CFT that is often only in the supergravity limit under control.

This talk reviews the results of~\cite{Blaszczyk:2009in}: the aim is to describe a new framework for phenomenologically attractive orbifolds, which in principle allows for a complete resolution (see e.g.~\cite{Lust:2006zh}). That is, which allows for the transition from the singular orbifold geometry to a smooth compactification by giving vacuum expectation values (VEVs) to blow-up modes (twisted states residing at the orbifold singularities). The key lies in the final step of gauge symmetry breaking from $\text{E}_8\times\text{E}_8$ to $\text{SU}(3)\times\text{SU}(2)\times\text{U}(1)_Y$ via $\text{SU}(5)$: this breaking is achieved by a freely acting Wilson line (WL) that is associated to a freely acting involution~\cite{Witten:1985xc}. It ensures that there is no flux in hypercharge direction, such that the hypercharge remains unbroken in the smooth limit. This talk focuses on the construction and geometrical interpretation of such involuted $\mathbbm{Z}_2\times\mathbbm{Z}_2$ orbifolds.

\section{A MSSM from the $\boldsymbol{\mathbbm{Z}_2\times\mathbbm{Z}_2}$ orbifold with additional involution}
The orbifold compactification is defined by the following steps. First, we choose a six-torus that is spanned by orthogonal vectors $e_\alpha$, $\alpha=1,\ldots,6$. Next, we identify a $\mathbbm{Z}_2\times\mathbbm{Z}_2$ symmetry of the torus generated by two rotations with respective phases $v_1 = (0,1/2,-1/2)$ and $v_2 = (-1/2,0,1/2)$ acting on the complex coordinates $z_i$, $i=1,2,3$, such that $\mathcal{N}=1$ supersymmetry is preserved in 4d. Finally, the geometrical action of the orbifold is embedded into the gauge degrees of freedom in terms of shifts $V_1$, $V_2$ and Wilson lines (WL) $W_\alpha$ that are chosen to be~\cite{Blaszczyk:2009in}
\begin{subequations}
\begin{eqnarray}
 V_1 & = & \left( \tfrac{5}{4},-\tfrac{3}{4},-\tfrac{7}{4}, \tfrac{1}{4}, \tfrac{1}{4},-\tfrac{3}{4},-\tfrac{3}{4}, \tfrac{1}{4}\right)\, 
           \left( 0, 1, 1, 0, 1, 0, 0,-1 \right)\;, \\
 V_2 & = & \left(-\tfrac{1}{2},-\tfrac{1}{2},-\tfrac{1}{2}, \tfrac{1}{2},-\tfrac{1}{2},-\tfrac{1}{2},-\tfrac{1}{2},-\tfrac{1}{2}\right)\, 
           \left(\tfrac{1}{2}, \tfrac{1}{2}, 0, 0, 0, 0, 0, 4 \right)\;, \\
W_2 & = & \left( \tfrac{5}{4}, \tfrac{1}{4}, \tfrac{3}{4},-\tfrac{1}{4},-\tfrac{1}{4}, \tfrac{3}{4}, \tfrac{3}{4}, \tfrac{3}{4} \right)\, 
          \left(-\tfrac{1}{4}, \tfrac{3}{4}, \tfrac{5}{4}, \tfrac{5}{4}, \tfrac{1}{4}, \tfrac{1}{4}, \tfrac{1}{4}, \tfrac{1}{4} \right)\;, \\
W_3 & = & \left(-\tfrac{3}{4},-\tfrac{1}{4}, \tfrac{1}{4}, \tfrac{7}{4},-\tfrac{1}{4},-\tfrac{1}{4},-\tfrac{1}{4},-\tfrac{1}{4} \right)\, 
          \left( \tfrac{1}{4}, \tfrac{1}{4}, \tfrac{1}{4}, \tfrac{5}{4},-\tfrac{3}{4}, \tfrac{1}{4},-\tfrac{3}{4}, \tfrac{1}{4} \right)\;, \\
W_5 & = & \left(-\tfrac{1}{2},-\tfrac{1}{2}, \tfrac{1}{2},-\tfrac{1}{2}, \tfrac{1}{2},-\tfrac{1}{2}, \tfrac{1}{2}, \tfrac{1}{2} \right)\, 
          \left( \tfrac{1}{2}, \tfrac{1}{2},   0,   0,   0,   0,-\tfrac{1}{2},-\tfrac{1}{2} \right) \;, \\
W_2 & = & W_4 ~=~ W_6\quad\text{and}\quad W_1 = \left(  0^{8} \right)\,\left(  0^{8} \right)\;.
\end{eqnarray}
\label{eqn:shiftsandWL}
\end{subequations}

\begin{figure}[t]
\begin{picture}(0,0)%
\includegraphics{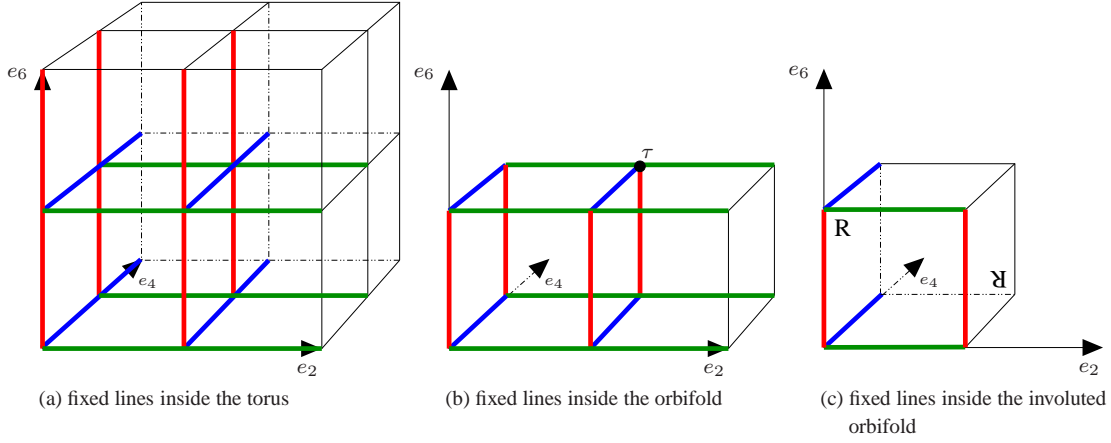}%
\end{picture}%
\setlength{\unitlength}{1948sp}%
\begingroup\makeatletter\ifx\SetFigFontNFSS\undefined%
\gdef\SetFigFontNFSS#1#2#3#4#5{%
  \reset@font\fontsize{#1}{#2pt}%
  \fontfamily{#3}\fontseries{#4}\fontshape{#5}%
  \selectfont}%
\fi\endgroup%
\begin{picture}(14256,5517)(931,-9976)
\put(14581,-9181){\makebox(0,0)[lb]{\smash{{\SetFigFontNFSS{8}{9.6}{\rmdefault}{\mddefault}{\updefault}$e_2$}}}}
\put(9021,-6461){\makebox(0,0)[lb]{\smash{{\SetFigFontNFSS{8}{9.6}{\rmdefault}{\mddefault}{\updefault}$\tau$}}}}
\put(4636,-9196){\makebox(0,0)[lb]{\smash{{\SetFigFontNFSS{8}{9.6}{\rmdefault}{\mddefault}{\updefault}$e_2$}}}}
\put(1351,-9601){\makebox(0,0)[lb]{\smash{{\SetFigFontNFSS{8}{9.6}{\rmdefault}{\mddefault}{\updefault}(a) fixed lines inside the torus}}}}
\put(6526,-9601){\makebox(0,0)[lb]{\smash{{\SetFigFontNFSS{8}{9.6}{\rmdefault}{\mddefault}{\updefault}(b) fixed lines inside the orbifold}}}}
\put(11656,-9961){\makebox(0,0)[lb]{\smash{{\SetFigFontNFSS{8}{9.6}{\rmdefault}{\mddefault}{\updefault}orbifold}}}}
\put(11296,-9601){\makebox(0,0)[lb]{\smash{{\SetFigFontNFSS{8}{9.6}{\rmdefault}{\mddefault}{\updefault}(c) fixed lines inside the involuted}}}}
\put(9811,-9196){\makebox(0,0)[lb]{\smash{{\SetFigFontNFSS{8}{9.6}{\rmdefault}{\mddefault}{\updefault}$e_2$}}}}
\put(2611,-8071){\makebox(0,0)[lb]{\smash{{\SetFigFontNFSS{6}{7.2}{\rmdefault}{\mddefault}{\updefault}$e_4$}}}}
\put(7786,-8071){\makebox(0,0)[lb]{\smash{{\SetFigFontNFSS{6}{7.2}{\rmdefault}{\mddefault}{\updefault}$e_4$}}}}
\put(946,-5416){\makebox(0,0)[lb]{\smash{{\SetFigFontNFSS{8}{9.6}{\rmdefault}{\mddefault}{\updefault}$e_6$}}}}
\put(6121,-5416){\makebox(0,0)[lb]{\smash{{\SetFigFontNFSS{8}{9.6}{\rmdefault}{\mddefault}{\updefault}$e_6$}}}}
\put(12556,-8056){\makebox(0,0)[lb]{\smash{{\SetFigFontNFSS{6}{7.2}{\rmdefault}{\mddefault}{\updefault}$e_4$}}}}
\put(10891,-5401){\makebox(0,0)[lb]{\smash{{\SetFigFontNFSS{8}{9.6}{\rmdefault}{\mddefault}{\updefault}$e_6$}}}}
\end{picture}%
\caption{Projection of the 6-dim. orbifold space on the 3-dim. subspace defined by $Re{(z_i)}=0$ for $i=1,2,3$. Part (a), (b) and (c) illustrate different steps of the compactification; the four/two (colored) fat lines parallel to $e_2$ are fixed lines of the $v_1 = (0,1/2,-1/2)$ sector, the ones parallel to $e_4$ belong to the $v_2 = (-1/2,0,1/2)$ sector and the ones parallel to $e_6$ are fixed lines of the $v_1+v_2 = (-1/2,1/2,0)$ sector. Strings from the twisted sectors are localized on such fixed lines. (a) depicts the fundamental domain of the torus. (b) shows the fundamental domain of the $\mathbbm{Z}_2\times\mathbbm{Z}_2$ orbifold, where $v_1$ has mapped the upper four boxes to the lower ones and $v_2$ has identified the two remaining boxes in the back with the ones in front resulting in one quarter of the fundamental domain of the torus. The freely acting involution $\tau$ is included as the vector pointing from the origin to the black dot. In (c) the $\tau$-action has been divided out, reducing the fundamental domain by another factor of two and identifying two fixed lines pairwise. Opposite faces of the box in (c) have to be identified after a $180^\circ$ rotation at the center of the face as exemplified by the letter R on the front and back.}
\label{fig:1}
\end{figure}

The orthogonal torus lattice and the equality of the WLs in the 2,4 and 6 direction allow for an additional orbifold identification according to the freely acting involution~\cite{Donagi:2004ht,Donagi:2008xy}
\begin{equation}
\label{eqn:tau}
 \tau ~=~ \frac{1}{2}\left(e_2 + e_4 + e_6\right)\;.
\end{equation}
The resulting orbifold is illustrated in fig.\ref{fig:1}. It turns out that there is a {\it freely acting Wilson line} $W$~\cite{Witten:1985xc,Hebecker:2003we,Hebecker:2004ce} along the $\tau$-direction given by the vector
\begin{equation}
 W~ =~ \frac{1}{2}\left( W_2 + W_4 + W_6 \right) ~=~ \frac{3}{2} W_2\;,
\end{equation}
which is of order 4 (i.e. $4W \in \Lambda_{\text{E}_8\times\text{E}_8}$) and hence stronger (in the sense of gauge symmetry breaking) than the original Wilson lines $W_\alpha$, $\alpha=2,4,6$. In our case, the $W_\alpha$ are chosen such that the freely acting WL $W$ breaks $\text{SU}(5)$ down to $G_\mathrm{SM} = \text{SU}(3)\times\text{SU}(2)\times\text{U}(1)_Y$. 

Furthermore, due to the additional $\tau$-direction, there are new (massive) winding modes with boundary conditions of the form $X(\sigma + 2\pi) = \pm X(\sigma) + n_0 \tau$ entering the orbifold partition function $Z$. Therefore, modular invariance of $Z$ imposes the following conditions on the freely acting Wilson line $W$
\begin{equation}
 4\,\left(n_\alpha\,W_\alpha+n_0\,W\right)^2  ~=~0\mod 2  \quad\forall~n_0,n_\alpha\in\mathbbm{Z}\;,
 \label{eqn:ModInvZ2free}
\end{equation}
in addition to the usual modular invariance constraints on shifts and Wilson lines, see e.g.~\cite{Ploger:2007iq}. One might speculate that conditions similar to eqn.~(\ref{eqn:ModInvZ2free}) have to be imposed also for smooth Calabi-Yau compactifications with freely acting involutions, e.g.~\cite{Bouchard:2005ag,Braun:2005nv}, in order to ensure that they are genuine string compactifications. However, for the orbifold model under investigation, defined in eqns.~(\ref{eqn:shiftsandWL}), these conditions are fulfilled.

\subsection{Massless spectrum and $B-L$}
The resulting model has Standard Model gauge group $G_\mathrm{SM}$ times a hidden sector $[\text{SU}(3)\times\text{SU}(2)\times\text{SU}(2)]_\mathrm{hid}$ and additional $\text{U}(1)$ factors, where one of them appears anomalous. Since the involution breaks $\text{SU}(5)$ to $G_\mathrm{SM}$, hypercharge originates from $\text{SU}(5)$ and is hence anomaly-free. Due to the same reason, the SM charged matter can only originate from $\text{SU}(5)$ representations (being $\boldsymbol{5}$, $\boldsymbol{\overline{5}}$ and $\boldsymbol{10}$ in this case) and consequently fractionally charged states are absent on the massless level~\cite{Schellekens:1989qb, Assel:2009xa}, c.f. the spectrum in table~\ref{tab:spectrum}. Similar to~\cite{Lebedev:2007hv} and using the techniques described in~\cite{Petersen:2009ip}, a non-standard $B-L$ generator
\begin{equation}
 \mathsf{t}_{B-L}~=~
 \left(-\frac{1}{2}, \frac{1}{2},-\frac{1}{2}, \frac{1}{2}, \frac{1}{2},-\frac{1}{6},-\frac{1}{6},-\frac{1}{6} \right)\,
 \left(\frac{13}{6},-\frac{7}{6}, \frac{5}{6}, \frac{5}{6}, \frac{5}{6}, \frac{5}{6},-\frac{1}{2},-\frac{1}{2} \right)
 \label{eqn:B-L} 
\end{equation}
can be identified that arises from both $\text{E}_8$ factors and hence allows for a breaking to matter parity $\mathbbm{Z}_2^\mathcal{M}$ (defined by $\exp{(2\pi\text{i}\frac{3}{2}q_{B-L})} = \pm 1$) induced by VEVs of states with $B-L$ charge $q_{B-L} = 2n/3$ for $n \in \mathbbm{Z}$. Therefore, $R$ parity violating couplings are forbidden and one can distinguish between the leptons $\ell$ and the down-type Higgs doublets $h$ and between $\overline{d}$ quarks and exotics $\overline{\delta}$. On the other hand dim. 5 proton decay operators remain problematic. To summarize, the spectrum contains three generations of quarks and leptons, four potential Higgs-pairs and some vector-like exotics, most of them in the hidden sector.

\begin{table}[t!]
\begin{center}
\begin{tabular}{|r|c|l|c|r|c|l|}
\hline
\# & representation & label & & \# & representation & label\\
\hline
$3$ & $(\boldsymbol{3}, \boldsymbol{2}; \boldsymbol{1}, \boldsymbol{1}, \boldsymbol{1})_{( \frac{1}{6}, \frac{1}{3})}$ & $q$ 
&& 
$3$ & $(\boldsymbol{\overline{3}}, \boldsymbol{1}; \boldsymbol{1}, \boldsymbol{1}, \boldsymbol{1})_{(-\frac{2}{3},-\frac{1}{3})}$ & $\overline{u}$   \\
$3$ & $(\boldsymbol{\overline{3}}, \boldsymbol{1}; \boldsymbol{1}, \boldsymbol{1}, \boldsymbol{1})_{( \frac{1}{3},-\frac{1}{3})}$ & $\overline{d}$ 
&&  
$3$ & $(\boldsymbol{1}, \boldsymbol{2}; \boldsymbol{1}, \boldsymbol{1}, \boldsymbol{1})_{(-\frac{1}{2}, -1)}$ &  $\ell$   \\
$3$ & $(\boldsymbol{1}, \boldsymbol{1}; \boldsymbol{1}, \boldsymbol{1}, \boldsymbol{1})_{( 1, 1)}$ &  $\overline{e}$ 
&&  
$33$& $(\boldsymbol{1}, \boldsymbol{1}; \boldsymbol{1}, \boldsymbol{1}, \boldsymbol{1})_{(0,a)}$ & $s$   \\
$4$ & $(\boldsymbol{1}, \boldsymbol{2}; \boldsymbol{1}, \boldsymbol{1}, \boldsymbol{1})_{(-\frac{1}{2}, 0)}$ & $h$ 
&& 
$4$ & $(\boldsymbol{1}, \boldsymbol{2}; \boldsymbol{1}, \boldsymbol{1}, \boldsymbol{1})_{( \frac{1}{2}, 0)}$ & $\overline{h}$    \\
$5$ & $(\boldsymbol{\overline{3}}, \boldsymbol{1}; \boldsymbol{1}, \boldsymbol{1}, \boldsymbol{1})_{( \frac{1}{3}, \frac{2}{3})}$ & $\overline{\delta}$ 
&& 
$5$ & $(\boldsymbol{3}, \boldsymbol{1};\boldsymbol{1}, \boldsymbol{1}, \boldsymbol{1})_{(-\frac{1}{3},-\frac{2}{3})}$ & $\delta$ \\
$5$ & $(\boldsymbol{1}, \boldsymbol{1}; \boldsymbol{3}, \boldsymbol{1}, \boldsymbol{1})_{(0,b)}$            &  $x$ 
&&  
$5$ & $(\boldsymbol{1}, \boldsymbol{1};\boldsymbol{\overline{3}}, \boldsymbol{1}, \boldsymbol{1})_{(0,-b)}$            & $\overline{x}$ \\
$6$ & $(\boldsymbol{1}, \boldsymbol{1}; \boldsymbol{1}, \boldsymbol{1}, \boldsymbol{2})_{(0, 0)}$            &  $y$ 
&&  $6$ & $( \boldsymbol{1}, \boldsymbol{1}; \boldsymbol{1}, \boldsymbol{2}, \boldsymbol{1})_{(0, 0)}$            & $z$    \\
\hline
\end{tabular}
\end{center}
\caption{Spectrum at the orbifold point. We show the representations w.r.t.\
$G_\mathrm{SM}\times\text{U}(1)_{B-L}\times[\text{SU}(3)\times\text{SU}(2)\times\text{SU}(2)]_\mathrm{hid}$ and
their multiplicities (\#) and labels. The $[\dots]_\mathrm{hid}$ groups stem
from the second $\text{E}_8$, and $a \in\{0, \pm 1, \pm 2, \pm 3\}$ and $b \in
\{-4/3, -1/3, 5/3\}$. The $B-L$ generator is given in
eqn.~(\ref{eqn:B-L}).}
\label{tab:spectrum}
\end{table}

\subsection{Turning on VEVs}

Driven by the Fayet-Iliopoulos $D$-term of the anomalous $\text{U}(1)$ some fields need to attain VEVs in order to preserve supersymmetry. Due to obvious phenomenological reasons we choose them to be singlets of $G_\mathrm{SM}\times\mathbbm{Z}_2^\mathcal{M}$, collectively denoted by $\phi_i$. Following the discussion in~\cite{Buchmuller:2006ik} we can construct $D=0$ solutions involving all the $\phi_i$ fields in terms of gauge invariant monomials $M \sim \phi_1^{n_1}\ldots\phi_N^{n_N}$ and $F=0$ solutions by considering the number of non-trivial $F$-terms and non-vanishing VEVs $\langle\phi_i\rangle$, being 44 for both in this example.  

After identifying the allowed terms in the superpotential (see appendix B of~\cite{Blaszczyk:2009in} for details), one finds effective mass matrices for the $\delta_i \overline{\delta}_j$ exotics and for the Higgs-pairs $h_i \overline{h}_j$, i.e.
\begin{equation}\label{eq:Mdelta}
\mathcal{M}^\delta_{ij} ~\sim~ \left(
\begin{array}{ccccc}
\phi^3 & s_1    &  \phi^3 & \phi^3 & \phi^3    \\
s_2    & \phi^3 &  \phi^5 & s_{16} & s_{20} \\
\phi^5 & \phi^3 &  \phi^5 & s_{26} & s_{31} \\
s_{28} & \phi^3 &  s_{19} & s_{10} & \phi^3    \\
s_{33} & \phi^3 &  s_{23} & \phi^3 & s_{10} \\
\end{array}
\right)\quad\text{and}\quad 
\mathcal{M}^h_{ij} \sim \left(
\begin{array}{cccc}
\phi^3 & s_3    & \phi^3 & \phi^3 \\
s_{15} & \phi^5 & s_{19} & s_{23} \\
\phi^3 & s_{26} & s_{10} & \phi^3 \\
\phi^3 & s_{31} & \phi^3 & s_{10} \\
\end{array}
\right)\;,
\end{equation}
where $\phi^n$ denotes a sum of (known) monomials in the VEVs $\langle\phi_i\rangle$ with $n$ being its lowest degree and $s_i$ denotes the explicit singlet label as listed in appendix A of~\cite{Blaszczyk:2009in}. Remarkably, all exotic triplets decouple at linear order in the VEVs. Unfortunately, the same is true for the Higgs-pairs in this configuration and hence one suffers under the stringy version of the $\mu$ problem. This problem can be solved in configurations with enhanced symmetries, which forbid the $\mu$ term at perturbative level~\cite{Kappl:2010pr}.

Finally, the Yukawa couplings of the quarks and of the charged leptons are of the form

\begin{equation}
\label{eqn:yuk}
\mathcal{M}^{\overline{u}} \sim \left(
\begin{array}{ccc}
\overline{h} \phi^4 & \overline{h} \phi^4 & 0 \\
\overline{h} \phi^4 & \overline{h} \phi^4 & 0 \\
0                   & 0                   & \overline{h}_1\\
\end{array}
\right) \quad\text{and}\quad 
\mathcal{M}^{\overline{d}} \sim \mathcal{M}^{\overline{e}} \sim \left(
\begin{array}{ccc}
0   & 0   & h_3 \\
0   & 0   & h_4 \\
h_3 & h_4 & 0   \\
\end{array}
\right)\;.
\end{equation}
The structure of these mass matrices reflects two interesting features: first, there is a $D_4$ symmetry of geometrical origin~\cite{Kobayashi:2006wq} (related to the vanishing WL $W_1=0$), where the first and the second generation form a $D_4$ doublet and the third generation a singlet, and secondly we find $\text{SU}(5)$ relations $\mathcal{M}^{\overline{d}} \sim \mathcal{M}^{\overline{e}}$ originating from the breaking of $\text{SU}(5)$ by the freely acting WL.

\begin{acknowledgement}
It is a pleasure to thank my collaborators M.~Blaszczyk, S.~Groot Nibbelink, M.~Ratz, F.~R\"uhle and M.~Trapletti for the successful collaboration and the organizers of the ``9th Hellenic School and Workshop on Elementary Particle Physics and Gravity'' for the great meeting. This work was supported by LMUExcellent.

\end{acknowledgement}

\bibliographystyle{fdp}

\begin{thebibliography}{[10]}

\bibitem{Dixon:1985jw}
 \textsc{L.\,J. Dixon},  \textsc{J.\,A. Harvey},  \textsc{C.~Vafa},  and
  \textsc{E.~Witten},
 \jr{Nucl. Phys.} \textbf{B261}, 678--686 (1985).


\bibitem{Dixon:1986jc}
 \textsc{L.\,J. Dixon},  \textsc{J.\,A. Harvey},  \textsc{C.~Vafa},  and
  \textsc{E.~Witten},
 \jr{Nucl. Phys.} \textbf{B274}, 285--314 (1986).


\bibitem{Ibanez:1986tp}
 \textsc{L.\,E. Ib{\'a}{\~n}ez},  \textsc{H.\,P. Nilles},  and
  \textsc{F.~Quevedo},
 \jr{Phys. Lett.} \textbf{B187}, 25--32 (1987).


\bibitem{Ibanez:1987sn}
 \textsc{L.\,E. Ib{\'a}{\~n}ez},  \textsc{J.\,E. Kim},  \textsc{H.\,P. Nilles},  and
  \textsc{F.~Quevedo},
 \jr{Phys. Lett.} \textbf{B191}, 282--286 (1987).


\bibitem{Kobayashi:2004ya}
 \textsc{T.~Kobayashi},  \textsc{S.~Raby},  and  \textsc{R.\,J. Zhang},
 \jr{Nucl. Phys.} \textbf{B704}, 3--55 (2005).


\bibitem{Buchmuller:2005jr}
 \textsc{W.~Buchm{\"u}ller},  \textsc{K.~Hamaguchi},  \textsc{O.~Lebedev},  and
  \textsc{M.~Ratz},
 \jr{Phys. Rev. Lett.} \textbf{96}, 121602 (2006).


\bibitem{Buchmuller:2006ik}
 \textsc{W.~Buchm{\"u}ller},  \textsc{K.~Hamaguchi},  \textsc{O.~Lebedev},  and
  \textsc{M.~Ratz},
 \jr{Nucl. Phys.} \textbf{B785}, 149--209 (2007).


\bibitem{Lebedev:2006kn}
 \textsc{O.~Lebedev} \etal{},
 \jr{Phys. Lett.} \textbf{B645}, 88--94 (2007).


\bibitem{Lebedev:2007hv}
 \textsc{O.~Lebedev} \etal{},
 \jr{Phys. Rev.} \textbf{D77}, 046013 (2008).


\bibitem{Nilles:2008gq}
 \textsc{H.\,P. Nilles},  \textsc{S.~Ramos-S{\'a}nchez},  \textsc{M.~Ratz},  and
  \textsc{P.\,K.\,S. Vaudrevange},
 \jr{Eur. Phys. J.} \textbf{C59}, 249--267 (2009).


\bibitem{Blaszczyk:2009in}
 \textsc{M.~Blaszczyk} \etal{},
 \jr{Phys. Lett.} \textbf{B683}, 340--348 (2010).


\bibitem{Lust:2006zh}
 \textsc{D.~L{\"u}st},  \textsc{S.~Reffert},  \textsc{E.~Scheidegger},  and
  \textsc{S.~Stieberger},
 \jr{Adv. Theor. Math. Phys.} \textbf{12}, 67--183 (2008).


\bibitem{Witten:1985xc}
 \textsc{E.~Witten},
 \jr{Nucl. Phys.} \textbf{B258}, 75 (1985).


\bibitem{Donagi:2004ht}
 \textsc{R.~Donagi} and  \textsc{A.\,E. Faraggi},
 \jr{Nucl. Phys.} \textbf{B694}, 187--205 (2004).


\bibitem{Donagi:2008xy}
 \textsc{R.~Donagi} and  \textsc{K.~Wendland},
 \jr{J. Geom. Phys.} \textbf{59}, 942--968 (2009).


\bibitem{Hebecker:2003we}
 \textsc{A.~Hebecker},
 \jr{JHEP} \textbf{01}, 047 (2004).


\bibitem{Hebecker:2004ce}
 \textsc{A.~Hebecker} and  \textsc{M.~Trapletti},
 \jr{Nucl. Phys.} \textbf{B713}, 173--203 (2005).


\bibitem{Ploger:2007iq}
 \textsc{F.~Pl{\"o}ger},  \textsc{S.~Ramos-S{\'a}nchez},  \textsc{M.~Ratz},  and
  \textsc{P.\,K.\,S. Vaudrevange},
 \jr{JHEP} \textbf{04}, 063 (2007).


\bibitem{Bouchard:2005ag}
 \textsc{V.~Bouchard} and  \textsc{R.~Donagi},
 \jr{Phys. Lett.} \textbf{B633}, 783--791 (2006).


\bibitem{Braun:2005nv}
 \textsc{V.~Braun},  \textsc{Y.\,H. He},  \textsc{B.\,A. Ovrut},  and
  \textsc{T.~Pantev},
 \jr{JHEP} \textbf{05}, 043 (2006).


\bibitem{Schellekens:1989qb}
 \textsc{A.\,N. Schellekens},
 \jr{Phys. Lett.} \textbf{B237}, 363 (1990).


\bibitem{Assel:2009xa}
 \textsc{B.~Assel},  \textsc{K.~Christodoulides},  \textsc{A.\,E. Faraggi},
  \textsc{C.~Kounnas},  and  \textsc{J.~Rizos},
 \jr{Phys. Lett.} \textbf{B683}, 306--313 (2010).


\bibitem{Petersen:2009ip}
 \textsc{B.~Petersen},  \textsc{M.~Ratz},  and  \textsc{R.~Schieren},
 \jr{JHEP} \textbf{08}, 111 (2009).


\bibitem{Kappl:2010pr}
 \textsc{R.~Kappl} \etal{}(2010).


\bibitem{Kobayashi:2006wq}
 \textsc{T.~Kobayashi},  \textsc{H.\,P. Nilles},  \textsc{F.~Pl{\"o}ger},
  \textsc{S.~Raby},  and  \textsc{M.~Ratz},
 \jr{Nucl. Phys.} \textbf{B768}, 135--156 (2007).


\end{thebibliography}

\providecommand{\WileyBibTextsc}{}
\let\textsc\WileyBibTextsc
\providecommand{\othercit}{}
\providecommand{\jr}[1]{#1}
\providecommand{\etal}{~et~al.}

\end{document}